\documentclass{article}
\usepackage{spconf,amsmath,graphicx,float}

\usepackage{enumitem}
\usepackage{amsmath,amssymb}
\setlist{nosep, leftmargin=14pt}

\usepackage{mwe}

\title{Enhancing Weakly Supervised Semantic Segmentation for Fibrosis via Controllable Image Generation}

\name{Zhiling Yue$^{1}$$^{*}$, Yingying Fang$^{2}$$^{*}$, Liutao Yang$^{2}$, Nikhil Baid$^{3}$, Simon Walsh$^{4}$, Guang Yang$^{2,4}$}

\address{$^{1}$ \small{Department of Surgery and Cancer, Imperial College London, London, UK} \\
$^{2}$ \small{Bioengineering Department and Imperial-X, Imperial College London, London, UK}\\
$^{3}$ \small{Computer Science Department, University College London, London, UK}\\
$^{4}$ \small{National Heart and Lung Institute, Imperial College London, London, UK}\\
\\
}

\begin{document}
%\ninept
%
\maketitle
\renewcommand{\thefootnote}{}
\footnotetext{$^*$Z. Yue and Y. Fang—Equal contribution.}

\begin{abstract}
Fibrotic Lung Disease (FLD) is a severe condition marked by lung stiffening and scarring, leading to respiratory decline. High-resolution computed tomography (HRCT) is critical for diagnosing and monitoring FLD; however, fibrosis appears as irregular, diffuse patterns with unclear boundaries, leading to high inter-observer variability and time-intensive manual annotation. To tackle this challenge, we propose DiffSeg, a novel weakly supervised semantic segmentation (WSSS) method that uses image-level annotations to generate pixel-level fibrosis segmentation, reducing the need for fine-grained manual labeling. Additionally, our DiffSeg incorporates a diffusion-based generative model to synthesize HRCT images with different levels of fibrosis from healthy slices, enabling the generation of the fibrosis-injected slices and their paired fibrosis location. Experiments indicate that our method significantly improves the accuracy of pseudo masks generated by existing WSSS methods, greatly reducing the complexity of manual labeling and enhancing the consistency of the generated masks.
\end{abstract}
%

% -------------------------------------------------------------------------
\section{Introduction}
\label{sec:intro}

\begin{figure}[h]
    \centering
    \includegraphics[width=\linewidth]{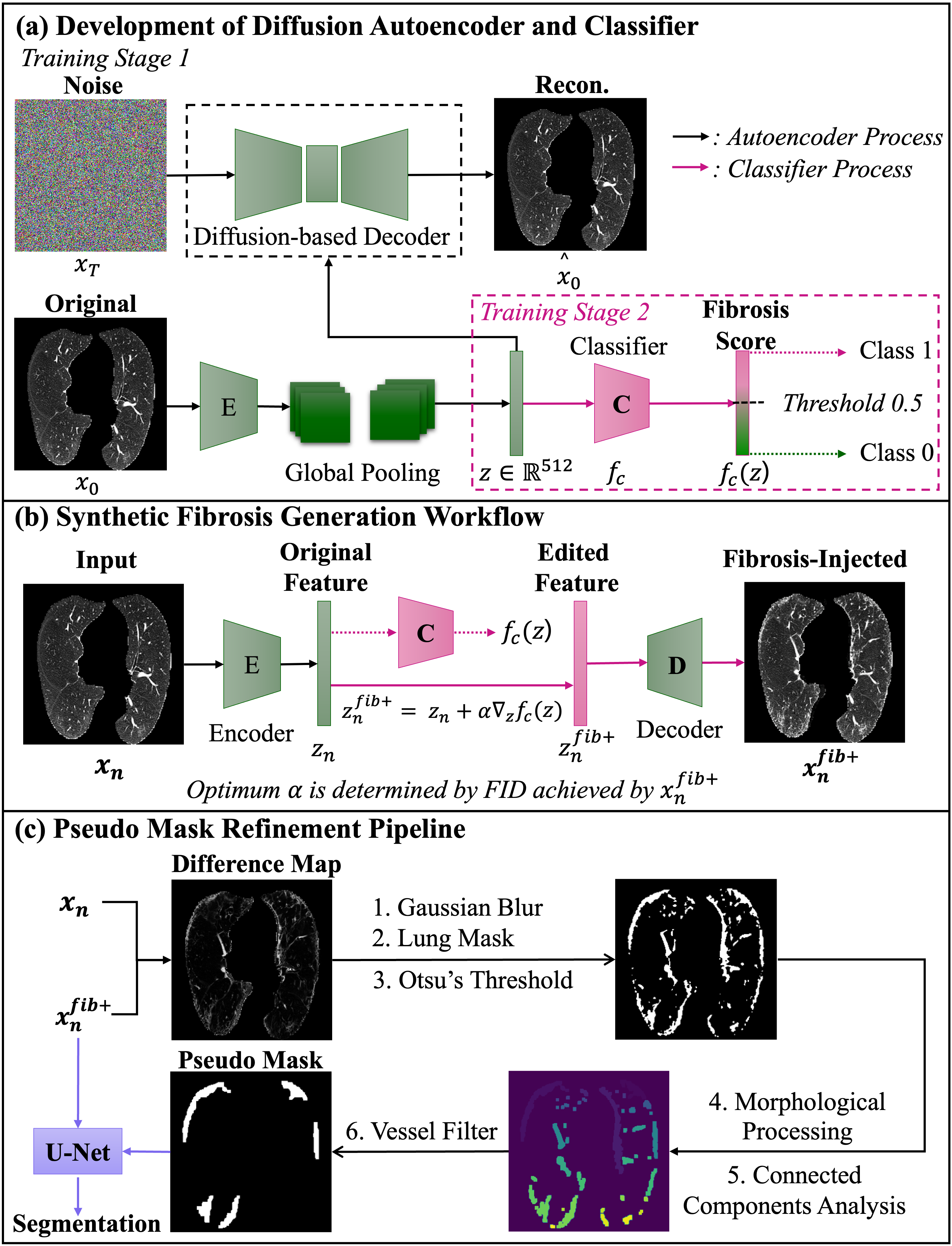}
    \caption{Framework of DiffSeg. (a) Pre-training of a diffusion-based autoencoder and classifier. (b) Slice-injected fibrosis generation workflow with trained models. (c) Pseudo mask generation and refinement process.}
    \label{fig:Framework_of_DiffSeg}
\end{figure}

Fibrotic Lung Disease (FLD) represents a group of severe conditions characterized by stiffening and scarring of the lungs, leading to progressive loss of respiratory function. Despite its significant health impact, FLD remains under-recognized; in 2021, it accounted for 1\% of all deaths in the United Kingdom \cite{FLD}, comparable to breast cancer mortality rates. Among FLD patients, a subset suffers from progressive pulmonary fibrosis (PPF), a rapidly advancing form with poor prognosis and a median survival time of just 2 to 5 years \cite{2-5years}. Early prediction of PPF is crucial to enable timely intervention.

High-resolution computed tomography (HRCT) is essential for diagnosing and monitoring FLD, with fibrosis extent on HRCT correlating strongly with mortality and serving as a prognostic marker \cite{HRCT}. However, fibrosis appears as irregular, overlapping patterns (e.g., honeycombing, reticulation and ground-glass opacity) that lack clear boundaries, leading to significant inter-observer variability \cite{inter-observer}. In recent years, computer-aided methods like DTA and INTACT \cite{data-driven,CAD} have emerged to automate segmentation and improve image analysis efficiency. Nevertheless, these methods rely heavily on extensive pixel-level annotations of high-dimensional CT volumes, and their performance is constrained by the availability of manual labels. 

Weakly Supervised Semantic Segmentation (WSSS) methods offer a promising alternative by using less detailed annotations and have shown strong performance in natural image segmentation, with potential applications emerging in medical image domain\cite{WSSS,WSSSMedical}. Recent image-level WSSS methods mainly rely on class activation maps, which tend to capture only the most discriminative regions and produce low-resolution maps, limiting segmentation quality \cite{DUPL}.

In this work, we introduce WSSS to the challenging task of fibrosis segmentation through a novel generative framework named Diffusion-Based Segmentation Model (DiffSeg). DiffSeg achieves pixel-level fibrosis annotation using only image-level labels (presence of fibrosis) provided by clinicians. It leverages a controllable latent space within a generative model to generate a fibrosis-injected HRCT slice from a fibrosis-free one, guided by a classifier pre-trained on image-level labels. This control ensures the synthesized and original slices are identical except for fibrosis patterns, enabling precise localization. The localized region is then refined into a pseudo mask to train final segmentation model. By combining controllable generative model and weak supervision, our approach enables WSSS for fine-grained,  medical segmentation tasks. This not only significantly reduces the manual labeling burden but also enhances the consistency of labels for boundary-ambiguous tasks.

\section{Methods}
\label{sec:methods}
Fig. \ref{fig:Framework_of_DiffSeg} provides an overview of the DiffSeg framework. As shown in Fig. \ref{fig:Framework_of_DiffSeg} (a) and (b), the encoder maps a fibrosis-free slice to a one-dimensional latent space, where a pre-trained classifier computes its fibrosis score and gradient. Next, we adjust the latent space by adding this gradient, scaled by a factor $\alpha$, and reconstruct the fibrosis-injected image from the modified latent feature. Finally, by computing the difference map and applying refinement pipeline, we obtain a fine-grained pseudo mask used to train a U-Net model for fibrosis segmentation.

\subsection{Diffusion-based Autoencoder}
\label{ssec:diffAE}
The Diffusion-based Autoencoder (DiffAE) is a generative model designed to encode images into a latent space, which serves as the conditioning input for diffusion model to reconstruct the image \cite{diffae}. In our weakly supervised generative framework, the DiffAE serves two primary functions, as shown in Fig. \ref{fig:Framework_of_DiffSeg}(b). First, it provides a controllable latent space for semantically meaningful manipulation guided by a pre-trained classifier. Second, it serves as generation condition, allowing the creation of edited images that align with human intentions and closely resemble real images, with only certain details changed.

In Fig. \ref{fig:Framework_of_DiffSeg}(a), DiffAE comprises an encoder (\( \mathbf{E}_\phi \)) and a decoder (\( \mathbf{D}_{\theta} \)), which are jointly trained using the loss function proposed in \cite{diffae}: 
\(\mathcal{L}_{(\phi, \theta)}=\mathbb{E}\left\|\boldsymbol{\epsilon}-\mathbf{D}_\theta\left({x}_t, t, \mathbf{E}_\phi({x}_0)\right)\right\|_1\).
The encoder down-samples the input image \( x_0 \) into a 512-channel feature map, transforming it into a one-dimensional latent vector \( z \in \mathbf{R}^{512} \). The decoder utilizes a conditional Denoising Diffusion Implicit Model (DDIM) \cite{DDIM,yyfDDIM} that takes \( z \) as a conditional input and perform the iterative denoising procedure as follows: 
\begin{equation}
\begin{aligned}
\hat{x}_{t-1} &= \frac{\sqrt{\alpha_{t-1}}}{\sqrt{\alpha_t}}  \left( \hat{x}_t - \sqrt{1 - \alpha_t} D_{\theta}(\hat{x}_t, t, z) \right)\\
&\quad + \sqrt{1 - \alpha_{t-1}} D_{\theta}(\hat{x}_t, t, z)\\
\end{aligned}
\end{equation}
where, \(\epsilon \sim \mathcal{N}(0, I), t = T, T-1, \dots, 1\)

\subsection{Fibrosis-Injection Manipulation}
\label{ssec:manipulation} 

To effectively generate paired fibrosis and it's paired mask, we proposed a \textit{Classifier-Guided, FID-Controlled} lesion generation process.
The role of classifier is to enable uncovering the semantic manipulation direction of fibrosis under the guidance of weak label without the need of a pixel-level label.

Specifically, the classifier is designed to be a one-layer architecture with encoded latent vector $z\in \mathbf{R}^{512}$ as input. To train the classifier in a weakly supervised manner, we designated slices marked as containing \textit{fibrosis} by clinicians as the positive class, while the remaining slices were assigned to the negative class. To enhance the classifier's ability to distinguish fibrosis more precisely, abnormal slices without fibrosis (e.g., lesion-containing but fibrosis-free) were also included in the negative class.

Apart from classifier guidance, the manipulation strength \( \alpha \) affects both the extent of generated fibrosis and the presence of artifacts.
To obtain the most accurate mask from the generated image, we aim to increase the fibrosis extent while minimizing artifacts and preserving realism. Specifically, we choose the optimal $\alpha$ which generates fibrosis-injected slices with the closest resemblance to real fibrotic slices, as measured by Fréchet Inception Distance (FID) \cite{FID}. Empirically, the optimal \( \alpha \) is determined to be 1.5.

\subsection{Pseudo Mask Refinement}
\label{sec:Pseudo Mask Generator}

Using paired synthetic fibrosis and lesion-free HRCT slices, we generate initial masks by calculating difference maps to capture injected fibrosis features. To achieve clearer and more human-interpretable masks, we introduce a Pseudo Mask Refinement pipeline, illustrated in Fig. \ref{fig:Framework_of_DiffSeg}(c). This process starts with a 5×5 Gaussian blur to reduce noise, followed by lung masking to remove outliers and Otsu’s thresholding to create binary masks. The masks are then undergo morphological operations to reduce noise by separating close regions. Subsequently, fibrosis patterns are identified by extracting contours and grouping connected components, retaining the five largest regions. A vessel filter is then applied to remove vessels related components for additional refinement. The resulting refined pseudo masks and synthetic fibrosis images are used to train a U-Net segmentation model \cite{U-Net} for the final segmentation task.

% -------------------------------------------------------------------------
\section{EXPERIMENTS}
\subsection{Datasets}

Our dataset is composed of two primary sources. The first is the OSIC Pulmonary Fibrosis Progression dataset \cite{OSIC}, consists of 52 HRCT scans with fibrosis annotations. The second source is our in-house ITAC dataset, which includes 566 HRCT scans from COVID-19 inpatients, used as negative cases (lesion-containing but fibrosis-free). %\cite{Covid-dataset}

For OSIC, image-level labels were assigned based on fibrosis presence: slices with fibrosis annotations were labeled as positive, while those without annotations were labeled as negative. The pixel-level annotations serve as the ground truth for segmentation results. For ITAC, we use all the slices as negative class. This process resulted in 12,625 fibrosis-positive slices and 12,625 fibrosis-free slices. To evaluate segmentation performance, 20\% of the fibrosis-positive slices were set aside for the final test.

\subsection{Experimental Details}

The diffusion model was trained on 200,000,000 samples, over 1,000 timesteps with a uniform timestep sampler on two A6000, each equipped with 40 GB of memory. The classifier was trained on image-level labels using the Cross Entropy Loss function, with the dataset split into training and testing sets at a 4:1 ratio. The model achieving the highest F1 score on the validation set (\(F_1 = 0.9328\)) was selected. For the U-Net segmentation model, 20\% of the synthetic fibrosis pairs were reserved for testing, with the remaining data split into training and validation sets at a 4:1 ratio. A five-fold cross-validation was used to ensure robustness, and the model with the highest Dice score on the validation set was selected.

Notably, for all training and validation experiments, only image-level labels are utilized, while pixel-level annotations are reserved exclusively for the testing phase. This setup ensures that our segmentation model operates under the sparsest supervision possible, as it never encounters the ground truth during training.

% -------------------------------------------------------------------------
 \section{RESULTS AND DISCUSSION}
\label{sec:Benchmarks}

\begin{figure}[tb]
    \centering
    \includegraphics[width=1.0\linewidth]{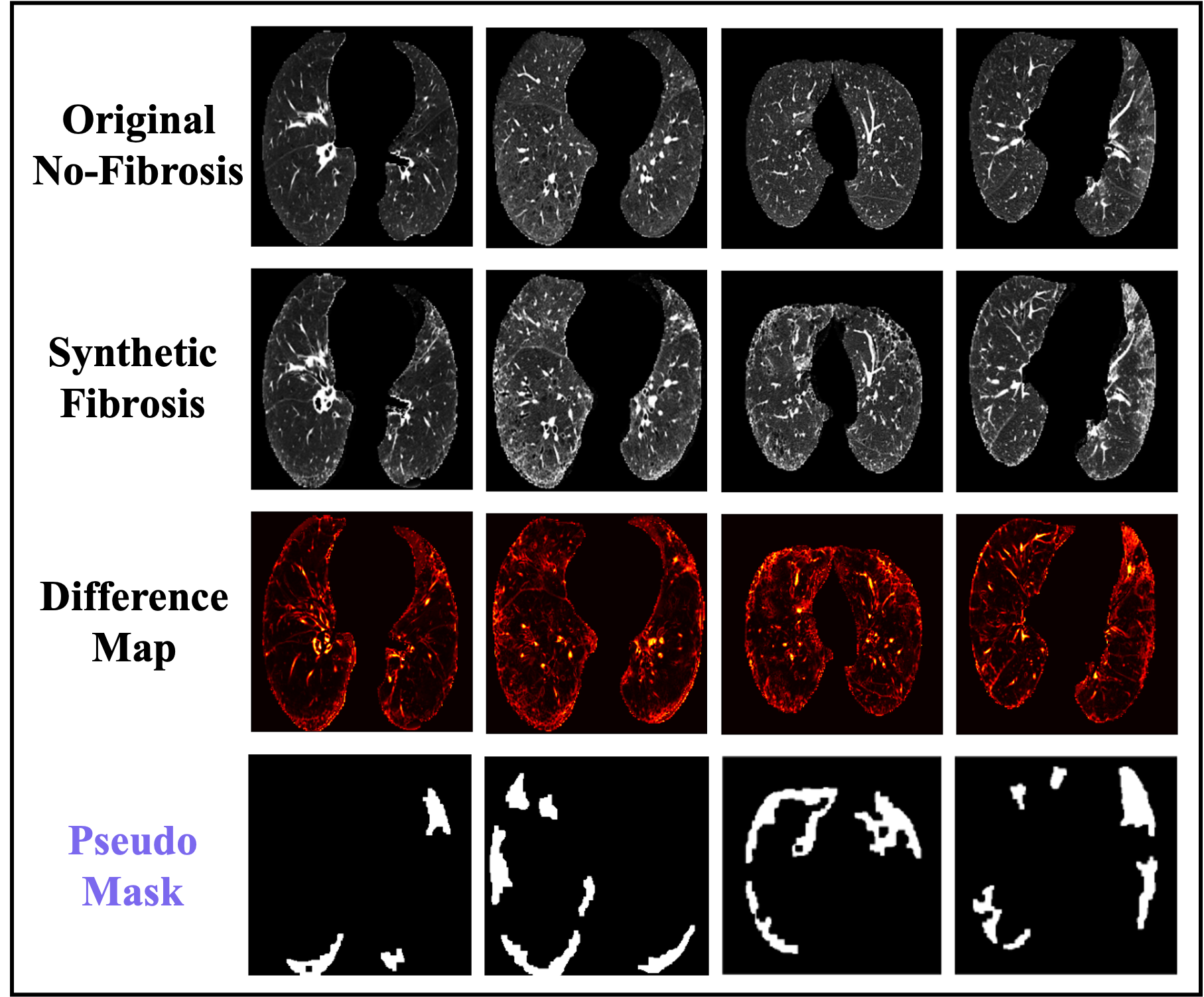}
    \caption{Pseudo mask refinement visualization.}
\label{fig:FID&PseudoMask}
\end{figure}

\begin{table}[ht]
\centering
\begin{tabular}{|l|l|l|l|}
\hline
\textbf{Method} & \textbf{Supervision} & \textbf{Backbone} & \textbf{Dice}$\uparrow$ \\ \hline
MedSAM   & Fine-box  & ViT-B      & 40.17\%                 \\ \hline
MedSAM   & Single-box    & ViT-B      & 26.31\%                 \\ \hline
DuPL     & Image-level & ViT-B      & 19.45\%                 \\ \hline
COIN     & Image-level & C-GAN      & 27.89\%                 \\ \hline
DiffSeg  & Image-level & Diffusion  & \textbf{61.75\%}                 \\ \hline
\end{tabular}
\caption{Semantic Segmentation Results. Fine-box denotes manually placing multiple fine boxes. Single-box denotes using one box to cover ROIs. Image-level labels only provide the presence of certain classes.}
\label{table results}
\end{table}

\begin{figure}[tb]
    \centering
    \includegraphics[width=1.0\linewidth]{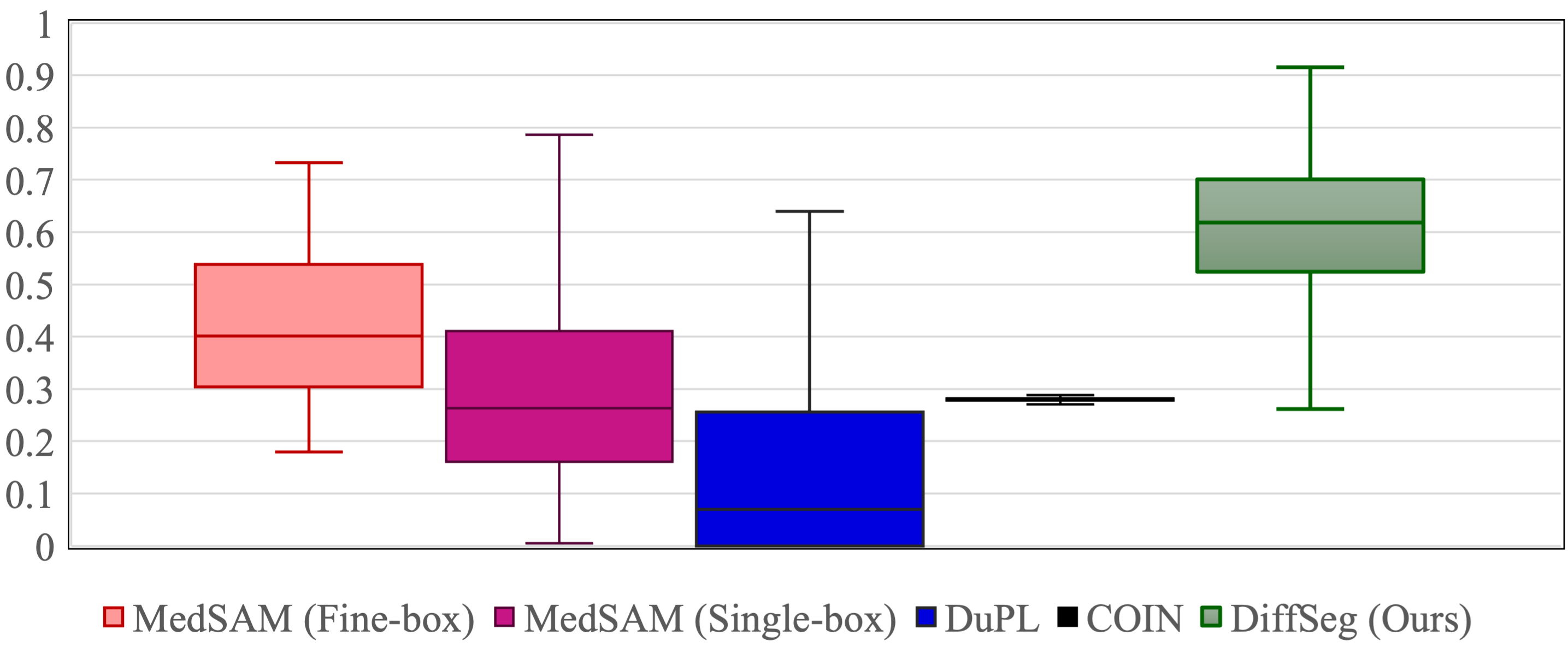}
    \vspace{-20pt} 
    \caption{Performance distribution of final segmentation task in terms of Dice score with interquartile range.}
    % \vspace{-10pt} 
\label{fig:BoxPlot}
\end{figure}

\begin{figure}[ht]
    \centering
    \includegraphics[width=1\linewidth]{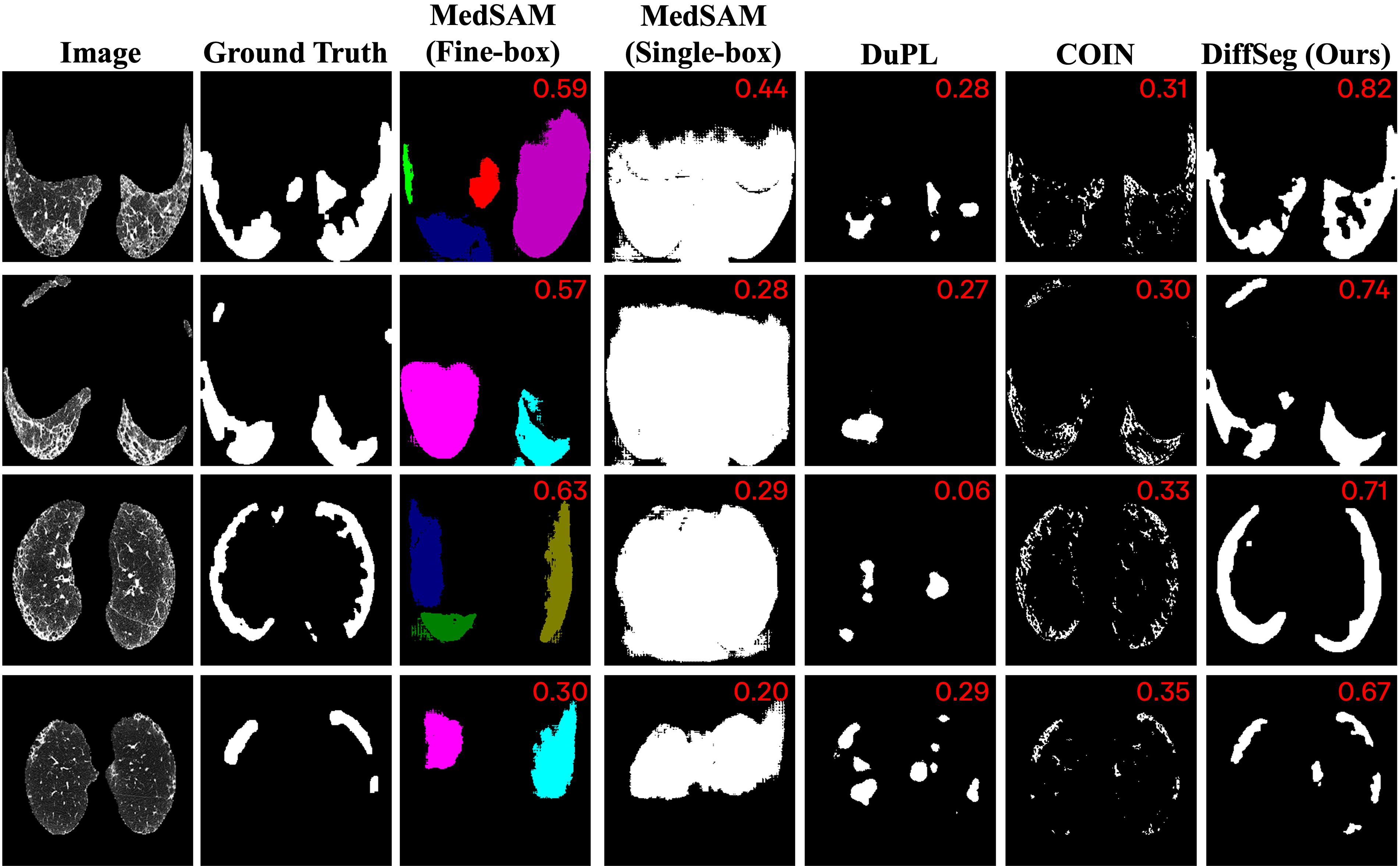}
    \caption{Segmentation visualization with Dice scores on top. Colors in MedSAM indicate several manually placed boxes.}
    \label{fig:seg visualization}
\end{figure}

\subsection{Synthetic Fibrosis Generation and Pseudo Mask}

As illustrated in Fig. \ref{fig:FID&PseudoMask}, the synthetic slices exhibit fibrosis-like patterns, primarily near the lung’s pleural surface (outer edges). These patterns include small, clustered cystic spaces that create a characteristic honeycombing appearance. Additionally, reticulation—a web-like network of intersecting linear opacities across the lung parenchyma—is visible, particularly in the fourth synthetic image. Both honeycombing and reticulation are hallmark patterns frequently observed in FLD.

The initial difference map effectively highlights the injected fibrosis patterns but includes some noise, such as vessels and lung boundaries. A pseudo mask is derived from this difference map and refined through a specialized pipeline, producing a cleaner, more accurate mask with reduced noise. This refined pseudo mask provides a more reliable reference for segmentation.

\subsection{Segmentation Performance on Real Fibrosis Slices}

To demonstrate the effectiveness of our approach, we compare DiffSeg with state-of-the-art WSSS methods that rely on image-level labels, including DuPL, which performs well on natural images, and COIN, a method with generative C-GAN model specialized for medical images \cite{DUPL,COIN}. To further highlight DiffSeg's strengths, we compare it with MedSAM, a large-scale foundation model for medical image segmentation, trained on over one million image-mask pairs across more than 100 medical datasets using bounding boxes for regions of interest (ROIs) \cite{MedSAM}. In our experiments, we first evaluated MedSAM with a single box around ROIs as recommended, but this lacked accuracy due to the diffuse, irregular nature of fibrosis. Multiple finely placed bounding boxes improved localization accuracy but required substantial annotation time. The Dice coefficient was used to assess segmentation performance across methods.

Table \ref{table results} and Fig. \ref{fig:BoxPlot} present the quantitative results for the final segmentation performance on fibrosis-positive slices. Overall, DiffSeg achieves the highest Dice score at \(61.75\%\) (interquartile range: 52.37-70.02\%), significantly surpassing recent WSSS methods (\(+42.3\%, +33.86\%\)) and outperforming MedSAM in both configurations with interactive box inputs (\(+35.44\%, +21.58\%\)). These results demonstrate that image-level supervision in DiffSeg is sufficient to achieve competitive segmentation performance, rivaling large-scale interactive models at segmenting challenging targets with indistinguishable boundaries.
 
In addition to quantitative comparisons, we visualize and compare segmentation masks from DuPL, COIN, MedSAM, and ground truth in Fig. \ref{fig:seg visualization}. DuPL struggled to accurately locate fibrosis patterns in the medical imaging domain. COIN identified some fibrosis-like patterns but was affected by noise near lung boundaries and produced unsmooth segmentation. MedSAM, using multiple finely placed bounding boxes, improved lesion pattern separation compared to single-box input, but still lacked precision in capturing the true shape of fibrosis patterns.

In contrast, our DiffSeg model produced segmentation that most closely resembled the ground truth with minimal noise, achieving the highest Dice scores and the most accurate fibrosis localization. Additionally, DiffSeg demonstrated the potential to identify more detailed and accurate target patterns (e.g., first and last rows of Fig. \ref{fig:seg visualization}). This advantage likely stems from the limitations of manual annotation, which often struggles to capture irregular, diffuse, and small patterns, leading to inter-observer variability. By comparison, DiffSeg's automated segmentation is more consistent and sensitive to intricate details, though further expert evaluation is required to fully validate these findings.

% -------------------------------------------------------------------------
\section{Conclusion}
\label{sec:CONCLUSION}

This study presents DiffSeg, a novel approach to advancing weakly supervised semantic segmentation for fibrosis detection in HRCT images. By relying on image-level labels rather than detailed pixel annotations, DiffSeg substantially reduces the annotation burden associated with fibrosis diagnosis. Leveraging a diffusion-based generative model, our method synthesizes fibrosis-injected images to create refined pseudo masks, enhancing segmentation accuracy. Experimental results show that DiffSeg outperforms state-of-the-art WSSS methods and MedSAM, achieving high accuracy with minimal annotation requirements. These findings highlight DiffSeg’s potential to streamline fibrosis monitoring and offer a practical, efficient alternative for indistinguishable boundary medical segmentation tasks.

\section{COMPLIANCE WITH ETHICAL STANDARDS}
The ITAC dataset used in this study  was obtained from the University Hospital of Parma, following the approval of the local Ethics Committee (code 934/2021/OSS/AOUPR - 11.01.2022). The OSIC dataset is openly accessible, with no ethical approval required as confirmed by its license.

% ------------------------------------------------------------------------- 
\bibliographystyle{IEEEbib}
\bibliography{Enhancing_WSSS_for_Fibrosis_via_Controllable_Image_Generation_Arxiv/refs}

\end{document}